\documentstyle[12pt]{article}

\def\bi{\bigskip}
\def\be{\begin{equation}}
\def\en{\end{equation}}
\def\bq{\begin{eqnarray}}
\def\eq{\end{eqnarray}}
\def\noi{\noindent}
\begin{document}

\begin{center}
{\bf Scalar meson masses and mixing angle in a $U(3) \times U(3)$ 
 Linear Sigma Model.}\\[1.5cm]
\end{center}
\vspace{1cm}
\begin{center}
{ \bf M. Napsuciale  }\\[1cm]
\vspace{.5cm}
{\it Instituto de Fisica, 
Universidad de Guanajuato.}\\
\vspace{.5cm}
{\it Apartado postal E-143,37150, Leon, Guanajuato, Mexico.}
\end{center}
\bi
\bi
\begin{center}
{\bf Abstract}
\end{center}

 Meson properties are considered within a $U(3)\times U(3)$ 
Linear Sigma Model(LSM). The importance of the $U_A(1)$-breaking term and the 
OZI rule violating term
in the generation 
of meson masses and mixing angles is stressed. 
The LSM parameters are fitted to the pseudoscalar meson spectrum thus giving
predictions for scalar meson properties. The model predicts a scalar meson 
$\bar q q$ nonet whose members are: $\{ \sigma(\approx 400), f_0(980), 
\kappa(\approx 900)$ and $a_0(980)\}$ resonances.  Scalar meson mixing 
angle (in the $\{|ns>, |s>\}$ basis) is predicted to be
$\phi_S\approx -14^\circ$. Therefore  the $f_0(980)$ is predominantly
 strange while 
the $\sigma(\approx 400)$ is mostly non-strange.  The model also gives a 
pseudoscalar mixing angle $\phi_P\approx 35^\circ$ which corresponds 
to $\theta_P \approx -19^\circ$ in the singlet-octet basis. Comparison with 
chiral perturbation theory  shows that $L_8$ is saturated by scalar mesons 
exchange. However, LSM predicts that $L_5$ is not saturated by scalar mesons 
exchange as assumed in the literature.

\noindent

\vspace{2cm}

\setlength{\baselineskip}{1\baselineskip}

\newpage

{\bf 1. Introduction}

\bi
Although we all believe in QCD as the theory describing strong interactions, 
its 
non-perturbative nature at low energies prevents its direct application  at
 the hadron level, thus being necessary to consider models for the strong 
interactions which exhibit the general features of QCD and be useful in the 
computation of hadron properties.
 
Concerning the light quark sector,the most important characteristic of QCD 
is the chiral symmetry exhibited by the QCD Lagrangian 
in the limit when the light quarks become massless. This has been 
the starting point for most of phenomenological approach to  low energy QCD. 
The whole chiral machinery have been
summarized in the so called Chiral Perturbation Theory (CHPT)[1] formalism 
(and its extensions [2,3,4]) which is still a very active field .

Another general approach which over the past few years has gained renewed 
interest is the $1/N_c$ expansion of QCD[5]. Concerning the meson sector 
this approach tell us that in the large $N_c$ limit, mesons are stable 
particles grouped in nonets, whose masses remain finite. Meson interactions 
are suppressed by $({1\over N_c})^{{n\over 2}-1}$ where $n$ denotes the 
number of mesons in the interacting vertex.

From the whole hadron spectrum the less-well-understood is the scalar sector.
In spite of the fact that the existence of the singlet scalar meson 
($\sigma $) was proposed long time ago, today we still do not have an 
unambiguous identification of this particle and the corresponding nonet of 
scalar mesons. The lowest lying
scalar mesons which have been firmly established are the I=0 ($f_0(980)$) and 
the I=1($a_0(980)$), but their  quark content is still controversial [6]. 
Among others, the  
argument against the considering of the $a_0(980)$ as the isovector $\bar q q$ 
resonance is its rather small coupling to two photons which is in disagreement
with naive quark model calculations [6,7]. Although a $\bar K K$ molecule 
picture calculation [7]
is closer to, it is still far from  experimental results[6,8]. Moreover , 
in the later calculations there are large 
uncertainties due to theoretical assumptions [7].

In addition to the above mentioned resonances we have  well-established 
scalar resonances above 1 GeV (e.g $K^*(1430)$)
and many others whose status is rather uncertain and have been conservatively 
labeled as 
($f_0(400-1200),f_0(1370), f_0(1500),a_0(1450)$ etc.) by the PDG[8].
The identification of the $q\bar q$ scalar resonances is further complicated 
by the possibility of the existence of 
non-$q\bar q$  states in the same energy region and with the same 
quantum numbers. From the ${1\over N_c}$ perspective 
some of these states ( glue-balls ) are also stable in the large $N_c$ limit. 

In the standard  CHPT machinery, 
scalar meson effects would appear at $O(p^4,p^6)$ due to the ``integrating 
out'' of these degrees of freedom. This is an acceptable  procedure whenever 
scalar 
meson masses lie above the  low energy region, typically above
 $\lambda_\chi$ (chiral symmetry breaking scale), and 
we are not interested in the 
scalar mesons phenomenology. However, it is well possible that a very broad 
and not so heavy scalar resonance ($\sigma $) exist [9,10]. In this case, 
these degrees of freedom should be 
explicitly included in the effective description of low energy QCD. 

The most 
promising possibility is to consider scalar degrees of freedom within  CHPT.
This theory is known to 
be a systematic expansion in powers of (small) momentum. However, predictive
power is substantially lost at $O(p^4)$ and beyond
due to the appearance 
of a large set of low energy constants (LEC) which have to be 
phenomenologically 
fixed . This problem can be circumvented by invoking  saturation by meson 
resonances [11]. Contributions of scalar meson resonances exchange to the 
CHPT LEC have 
been considered in the literature [11]. Here again, the problem 
of the scalar mesons arise as we still don't have confident experimental 
information on these resonances and is not possible to make reliable 
estimations of the contribution of scalar resonances exchange to CHPT low 
energy constants. It has been assumed e.g. that scalar meson resonances 
exchange saturate $L_5$ and $L_8$ [11].

\noi
Extension  of CHPT beyond the very low energy domain is problematic. 
The heavy 
field approach is necessary for spin 1/2 fermions  (and vector mesons) which 
forces 
us to consider a double expansion and to the introduction of spin 3/2 degrees 
of freedom [2]. Again, introduction of spin 3/2 resonances forces us  to 
consider a new (triple) expansion(``small 
scale'' expansion[3]) and the theory rely upon a spin 3/2 
field theory which for decades has been known to suffer from serious yet 
unresolved inconsistencies{\footnote {it has been  argued in [12]  that only 
the leading order term in the 1/m expansion of spin ${3\over 2}$ fields could 
be theoretically consistent. Explicit calculations [3] and a Hamiltonian 
quantization of the theory (see Eq.(28) of Ref.[13]) gives support to this 
argument.}}.
On the other side, it is well known that 
``golden plated'' (i.e. low energy parameters free) predictions of 
CHPT at $O(p^4)$ (e.g. $\gamma \gamma \to \pi^0\pi^0$ [14])  fail in the 
description of data at threshold were it is supposed to work well. Perhaps 
this is an indication that 
we are already missing something which well it could be the contributions of 
a not so heavy scalar meson. From the CHPT perspective this is equivalent 
to say that $O(p^6)$ contributions to that
process are saturated by  scalar mesons exchange.

Another possibility for the description of scalar mesons phenomenology is
 the linear realization of chiral symmetry. However, although a pure-meson 
LSM is a renormalizable model, it is not clear how a 
systematic expansion can be carried out as there is no obvious small parameter.
Nevertheless, it is symptomatic that one-loop corrections e.g. to meson 
masses have 
been found by Chan and Haymaker to be small[15]($\simeq 10\%$). This raises 
the possibility of a weak
regime in the effective LSM description of low energy QCD at least in 
the meson sector. In this concern, it is quite remarkable that in a 
$SU(2)\times SU(2)$ LSM the derivative 
coupling of pions (characteristic of non-linear realizations) is recovered  
in the LSM after delicate (``miraculous''[16]) cancellations when we expand 
in inverse powers of the center of mass energy. In this sense a LSM gives 
the same results (at leading order) as CHPT in the very low energy 
regime [16].

From the ${1\over N_c}$ perspective, a  LSM description of low energy 
($\bar q q$) meson-meson 
interactions can be accurate whenever other ${1\over N_c} \to 0 $ surviving 
non-$\bar q q$ states
be much more massive than the LSM degrees of freedom. 
From lattice calculations and theoretical 
estimates [17] the lowest lying scalar glue-ball 
has a mass of $\simeq 1.5 GeV$ or above, hence a  LSM 
can be naively expected to be  a good approximation below $1 GeV$.

Recently, there has been renewed interest in the linear realization of chiral 
symmetry 
[9,18,19,20]. Some linkage between a $SU(2)\times SU(2)$ quark-level LSM 
and non-perturbative QCD 
have been established [18] and double-counting ambiguities have  been 
addressed [19].
On the other hand, signals of scalar 
($\sigma (650)$ and $\kappa (900)$) resonances  in reanalysis of 
$\pi \pi $-$K K$ scattering have been argued[20] to be reproduced  within a 
pure-meson LSM in its $U(3)\times U(3)$ version[21,22].

\noi 
The pure-meson $U(3)\times U(3)$ LSM  has also been used by Tornqvist as a 
field theoretical 
framework where to interpret the results arising from his unitarized quark 
model [9]. However in the Ishida reanalysis of data[20]
the $I=1/2$ scalar resonance turns out to have a mass around $900 MeV$ 
whereas Tornqvist identifies this resonance with the $K^*(1430)$ 
and some signal below the $K-\pi$ threshold [9].

\noi
Although the $U(3)\times U(3)$ LSM model has previously been studied in 
the literature [15,21,22], we consider that a complete analysis of its 
phenomenological implications  is still missing. This is particularly important
for scalar mesons since available experimental information on these particles
 is unclear as can be seen from 
the last three versions of the RPP. Furthermore, the theory ( when written in 
terms of the 
appropriate fields) allow us to transparently identify the corresponding 
quark content of mesons which is particularly useful in the identification 
of $\bar q q$ scalar mesons. 

In this work we explore  predictions for scalar 
mesons properties arising from a $U(3)\times U(3)$ LSM with $U_A(1)$ breaking.
 This model has 
already been considered by t'Hooft [23], in its $U(2)\times U(2)$ version, in 
connection with the explanation of the $U_A(1)$-problem by instanton 
contributions.
 We begin by carrying  out a detailed 
analysis of the $U(3)\times U(3)$ LSM predictions for meson  masses and 
mixing angles. We shall be concerned with the meson sector only putting 
fermions aside for a while as they are not relevant for the 
physical quantities studied here.
We consider a $U(3)\times U(3)$ LSM in order to 
phenomenologically incorporate the effects of the $U_A(1)$ symmetry breaking.
 These effects are well known in the pseudoscalar sector but, as we shall see 
below, they give the scalar sector unexpected properties. The $U_A(1)$ breaking
 effects in the scalar sector are explicitly exhibited and turn out to be 
crucial in the understanding of scalar mesons. 
 
\noi
In order to make contact with the 
quark content of scalar mesons we start working in the usual singlet-octet 
basis but we switch later to the strange-non-strange basis[24].
Whenever possible, we use  
pseudoscalar mesons experimental information in order to fix the parameters 
of the model in such 
a way that we can make predictions for scalar mesons. It turns out 
that one of the parameters ($\lambda^\prime$) cannot be fixed from 
pseudoscalar spectrum as only the scalar mixed sector
depends directly on it. Hence, we choose to 
analyze the mixed scalar meson masses and mixing angle dependence 
on $\lambda^\prime$. 
This parameter has been suggested by Tornqvist to be small 
($\lambda^\prime$ used in this work 
corresponds to $4\lambda^\prime_T$ in Ref.[9]), and bosonization of a 
Nambu-Jona-Lasinio (NJL) model extended with a t'Hooft like interaction [25] 
predicts $\lambda^\prime \approx {\cal O}(1/N^3_c)$. As the author of [25] 
points out, this result is consistent with but not necessarily required by 
QCD. We include this term here as it is compatible with chiral symmetry and 
renormalizability in the usual sense. 

\noi
In spite of some sensitivity of scalar meson masses to the $SU(3)$ symmetry 
breaking, the model indicates that is likely to 
identify the scalar nonet with the 
$\{\sigma(\approx 400),f_0(980),\kappa(900)$ and $ a_0(980)\}$ resonances 
which 
according to the above discussion leads to a scenario where the (broad) sigma
 meson would give dominant contributions in the very low energy domain of QCD.
We obtain a scalar mixing angle $\phi_S \approx -14^\circ$.
The pseudoscalar mesons mixing angle predicted by this model is 
$\phi_P\approx 35^\circ$ which corresponds to $\theta \approx -19^\circ$ in 
the usual singlet-octet basis. 

\noi
We also study scalar mesons exchange contributions to the ${\cal O}(p^4)$ CHPT 
low energy constants. Expectations for the saturation of $L_8$ are corroborated
but LSM predicts that $L_5$ is not saturated by meson resonances exchange. 

\bi

{\bf 2. $U(3) \times U(3)$ Linear Sigma Model.} 

The $U(3)\times U(3)$ LSM was originally proposed by Levy[19] in a quark-meson 
version. 
The meson sector of this theory was also considered by Gasiorowicz 
and Geffen [22].

\noi
The  $U(3) \times U(3)$ symmetric Lagrangian  is written in terms of a 
nonet of scalar $(\sigma_i)$ and a nonet of pseudo-scalar
$(P_i)$ fields. A nonet is represented by the matrix: 
$$F\equiv \frac{1}{\sqrt{2}} \lambda_i f_i \qquad\qquad i=0,\ldots  8.$$
\noi where $\lambda_i  (i=1,\ldots , 8)$ are the Gell-Mann matrices, $\lambda_0
\equiv\sqrt{\frac{2}{3}} {\bf 1}$ and $f_i$ denotes a generic field. We will 
use $\bigl<\;\bigr>$ throughout this paper to denote trace over the $U(3)$ 
structure.

\noi
The Lagrangian is given by
\be
{\cal L}= {\cal L}_{sym}  +{\cal L}_{SB}
\en
where ${\cal L}_{sym}$ denotes the $U(3)\times U(3)$ symmetric Lagrangian:

$${\cal L}_{sym}=\bigl<\frac{1}{2}(\partial_\mu M)
(\partial^\mu M^{\dagger}) \bigr>  -\frac{\mu^2}{2} 
X (\sigma ,P)-\frac{\lambda}{4} Y(\sigma ,P)-
\frac{\lambda^\prime}{4} X^2 (\sigma , P)$$
\noi with $M=\sigma + i P$, and $X,Y$ stand for the $U(3)\times U(3)$ 
chirally symmetric terms:
$$
X(\sigma ,P)=\bigl< M M^{\dagger} \bigr> $$
$$Y(\sigma,P)= \bigl<(M M^{\dagger})^2\bigr>=\bigl<(\sigma^2+P^2)^2+2(
\sigma^2P^2-(\sigma P)^2)\bigr>.
$$
The chiral symmetry breaking terms are given by 
\bq
{\cal L}_{SB} =  \bigl< c\sigma \bigr> - \beta Z(\sigma,P)
\eq
where
\bq
Z(\sigma ,P) &=& \{ det (M)+ det (M^{\dagger})\} \\ \nonumber
&=& {1\over 3}[\bigl<\sigma \bigr>(\bigl<\sigma\bigr>^2-3\bigl<P\bigr>^2)-
3 \bigl<\sigma \bigr>\bigl<\sigma^2 -P^2\bigr> 
+ 6\bigl<P\bigr>\bigl<\sigma P\bigr> \\
& & + 2 \bigl<\sigma (\sigma^2 -3 P^2)\bigr>] \nonumber
\eq
\noi and $c\equiv \frac{1}{\sqrt{2}} \lambda_i c_i$, with $c_i$ constant.
The most general form of $c$ which preserves isospin and gives PCAC is such 
that the only non-vanishing coefficients are $c_0$ and $c_8$. The former 
gives, by hand, the pseudo-scalar nonet a common mass, while the later 
breaks the $SU(3)$ symmetry down to isospin. These parameters can be  
related to 
quark masses in QCD.
The $Z$ term in Eq.(2) is $U(3)_V\times SU(3)_A$ symmetric and is reminiscent
 of the quantum effects in QCD breaking the $U_A(1)$ symmetry.

 Performing the customary calculations we obtain
for the divergences of the vector and axial-vector currents:
\bq
\partial_\mu A^\mu\,_j &=& (c_0 ~d_{ojk} + ~c_8 ~d_{8jk}) P_k  \\
\partial_\mu V^\mu\,_j &=& c_8 f_{8jk} \sigma_k . \nonumber 
\eq
The choice (2) for the breaking term gives PCAC in the usual way
\bq
\partial_\mu A^\mu\,_{\vec\pi} &=& (\sqrt{\frac{2}{3}}c_0 + \sqrt{\frac{1}{3}}
c_8) \vec\pi \equiv m^2_\pi f_\pi \vec\pi \\
\partial_\mu A^\mu\,_k &=& (\sqrt{\frac{2}{3}} c_0 -\frac{1}{2} \sqrt{\frac{1}{3}} c_8) K \equiv m^2_K f_K K \nonumber 
\eq

\noi
The  linear $\sigma$  term in Eq. (2) induces $\sigma$-vacuum transitions 
which give to  $\sigma$  fields  a  non-zero vacuum expectation value 
(hereafter denoted by $\{~\}$). Linear terms can be eliminated  from  the  
theory  by  performing a shift to a new 
scalar  field  $S=\sigma -V$ such that $\{S\}=0$,  where  $V\equiv\{\sigma\}$.
This shift generates new three-meson interactions and mass terms.    

\noi
For the sake of simplicity lets write $V= Diag (a,a,b)$ where $a,b$ are 
related to $\{\sigma \}$ through

\bq
a&=&\frac{1}{\sqrt{3}}\{\sigma_0\}+\frac{1}{\sqrt{6}}\{\sigma_8 \} \\
b&=&\frac{1}{\sqrt{3}}\{\sigma_0\}-\frac{2}{\sqrt{6}}\{\sigma_8\}\nonumber
\eq

\noi
It is convenient to write the shifted meson Lagrangian as:
\bq
{\cal L}_{MM} (\sigma , P) = \sum^4_{n=0} {\cal L}_n 
\eq
\noi where terms  containing products of $n$ fields are collected in 
${\cal L}_n$. Explicitly
\bq
{\cal L}_0 &=&-\frac{\mu^2}{2} (2a^2+b^2)- \frac{\lambda^\prime}{4} (2a^2+b^2)^2-\frac{\lambda}{4} (2a^4+b^4)-2\beta a^2b+ \bigl<CV\bigr> \nonumber \\
{\cal L}_1 &=& -(\mu^2+\lambda^\prime (2a^2+b^2))\bigl<SV\bigr>-\lambda ((a^2+ab+b^2)\bigl<SV\bigr>-ab(a+b)\bigl<S\bigr>) \nonumber \\
&~& -2\beta(a(a+b)\bigl<S\bigr>-a\bigl<SV\bigr>)+\bigl<CS\bigr> \nonumber \\
{\cal L}_2 &=& -\frac{\mu^2}{2}X(S)-\lambda^\prime (\bigl<SV\bigr>^2+\frac{1}{2} (2a^2+b^2)\bigl<S^2+ P^2\bigr>) \\ 
&~& -\lambda \bigl<\bigg((a+b)V-ab\bigg)(S^2+P^2)+ \frac{1}{2} (VS)^2 -\frac{1}{2} (VP)^2\bigr> \nonumber \\
&~& -\beta[(2a+b)(\bigl<S\bigr>^2-\bigl<P\bigr>^2-\bigl<S^2-P^2\bigr>)-2\bigl<S\bigr>\bigl<VS\bigr> +\nonumber \\
&~&2\bigl<P\bigr>\bigl<VP\bigr>+2\bigl<V(S^2-P^2)\bigr>] \nonumber \\
{\cal L}_3 &=&-\beta Z(S)-\lambda^\prime \bigl<SV\bigr>\bigl<S^2+P^2\bigr>-\lambda \bigl<V(S^3+P^2S+SP^2-PSP)\bigr> \nonumber \\
{\cal L}_4 &=&-\frac{\lambda}{4} Y(S,P) -\frac{\lambda^\prime}{4} X^2 (S,P) \nonumber
\eq
The condition ${\cal L}_1=0$ gives:
\bq
\sqrt{\frac{2}{3}} c_0 + \sqrt{\frac{1}{3}} c_8 &=& \sqrt{2} a (\xi
+2\beta b+\lambda a^2) \nonumber \\
\mbox {}\\
\sqrt{\frac{2}{3}} c_0-\frac{1}{2}\sqrt{\frac{1}{3}} c_8 &=& \frac{1}{\sqrt{2}}
(a+b)\bigg(\xi+2\beta a+\lambda (a^2-ab+b^2)\bigg) \nonumber
\eq
where, for convenience, we defined $\xi \equiv \mu^2 +\lambda^\prime(2a^2+b^2)$.
\bi

{\bf 3.  Meson masses and mixing angles.}

\bi

Meson masses are given by ${\cal L}_2$. A straightforward calculation 
reveals the following masses for the non-mixed sectors.
\bq
m^2_\pi &=&\xi+2\beta b+\lambda a^2 \cr
m^2_K &=& \xi+2\beta a+\lambda (a^2-ab+b^2) \cr
m^2_a &=& \xi-2\beta b +3\lambda a^2 \\
m^2_\kappa &=&\xi-2\beta a+\lambda (a^2+ab+b^2) , \nonumber
\eq
\noi where we have denoted $a$ and $\kappa$ the scalar mesons anologous to
 $\pi$ and $K$.
\noi 
For the mixed sectors we obtain the following Lagrangian.
\bq
{\cal L}^P_2 =-\frac{1}{2} (m^2_{0P} P^2_0+ m^2_{8P} P^2_8 + 
2m^2_{08P} P_0 P_8) \\
{\cal L}^S_2 =-\frac{1}{2} (m^2_{0S} S^2_0+ m^2_{8S} S^2_8 + 
2m^2_{08S} S_0 S_8) \nonumber
\eq
\noi where
\bq
m^2_{0P} &=& \xi+ {1\over 3} [\lambda (2 a^2 + b^2) - 4
\beta (2a+b)] \nonumber \\
m^2_{8P} &=& \xi+ {1\over 3} [\lambda (a^2 +2 b^2) + 2\beta (4a - b)] \\
m^2_{08P} &=& \frac{\sqrt{2}}{3} (a-b)\bigg(\lambda (a+b) + 2\beta \bigg)
\nonumber \\
m^2_{0S} &=&\xi+\frac{1}{3} [4\beta(2a+b)+3\lambda (2a^2+b^2)+
2\lambda^\prime (2a+b)^2] 
\nonumber \\
m^2_{8S} &=& \xi+\frac{1}{3} [-2\beta (4a-b)+3\lambda (a^2+2b^2)+
4\lambda^\prime (a-b)^2] \nonumber \\
m^2_{08S}&=&\frac{\sqrt{2}}{3}(a-b)[-2\beta+3\lambda(a+b)+
2\lambda^\prime(2a+b)]\nonumber
\eq
In the case of a $SU(3)$ symmetric vacuum 
($a=b$ or equivalently $\{\sigma_8\}=0$), 
the pseudoscalar octet has a common squared mass which differs from
the pseudoscalar singlet squared mass by a term proportional to 
$\beta$ (the $U_A(1)$-breaking term ). Keeping $a\neq b$ amount to 
simultaneously consider $U_A(1)$-breaking and $SU(3)$ breaking contributions 
to the $\eta -\eta^\prime$ splitting. Similar results are obtained for the 
scalar nonet but also $\lambda^\prime$ contributes to the octet-singlet 
squared mass difference in this case. In order to exhibit the mass pattern in 
the general case 
of chiral symmetry broken down to isospin, and in order to asses the relevant
aspects of the theory it is instructive to
shift to new fields in the mixed sector. This change of basis explicitly 
shows the role played by the $U_A(1)$-breaking  term in the generation of 
scalar and pseudoscalar meson masses. 
Also, quark content is transparent and the role played by the different terms 
in the generation of masses, mixing, and violation to OZI rule, is 
obvious. 
Switching to the new fields ($\{|s>, |ns>\}$ basis [24]) defined by:
\bq
 P_{ns}=\sqrt{1\over 3} P_8  + \sqrt{2\over 3}P_0  \\
 P_{s} =-\sqrt{2\over 3 } P_8  + \sqrt{1\over 3} P_0  \nonumber
\eq 
and similar relations for the scalar mixed fields, we obtain an analogous 
Lagrangian to Eq. (11) where
\bq
m^2_{P_{ns}} &=&\xi-2\beta b + \lambda a^2 \cr
m^2_{P_s} &=&\xi + \lambda b^2 \cr
m^2_{P_{s-ns}} &=& -2\sqrt{2}\beta a \\
m^2_{S_{ns}} &=&\xi+2\beta b + 3\lambda a^2 +4\lambda^\prime a^2 \cr
m^2_{S_s} &=&\xi + 3\lambda b^2 + 2\lambda^\prime b^2 \cr
m^2_{S_{s-ns}} &=& 2\sqrt{2}(\beta+\lambda^\prime b)a \nonumber
\eq

\noi
The vacuum expectation values of the $s-ns$ scalar fields  are simply 
related to the parameters $a, b$ as: $\{\sigma_{ns}\}=\sqrt{2}a , 
\{\sigma_s\}=b$.

\noi
From  the above relations is clear  that in the $\beta = 0 $ case, the 
pseudoscalar sector gets diagonalized and even in the general case 
when chiral symmetry is  broken down to 
isospin ($ a\not=b$), the $U_A(1)$-breaking term 
$Z$ is required in order to break the $\eta_{ns}-\pi$ degeneracy [22]. In 
the scalar sector, however, we 
still have a non-zero mixing when $\beta = 0$ due to the $\lambda^\prime$ term.

It is interesting to analyze, step by step, the results so far obtained
 from a $ 1/N_c$ perspective.
According to $1\over N_c$ counting rules [5], OZI rule allowed 
three(four)-meson interactions 
should be 
${1\over \sqrt{N_c}}$($1\over N_c $) suppressed. OZI rule violating terms, like
 $\beta$ and $\lambda^\prime$ in Eq.(1), are further suppressed by higher
 powers of $1\over N_c$. As shown in [25], $\beta \approx {\cal O}
({1\over N^{3/2}_c})$ while $\lambda^\prime$ is even more suppressed.  Thus, 
based on ${1\over N_c}$ counting rules, 
we would expect $\beta$ and $\lambda^\prime$ contributions to physical 
quantities to be small as compared with $\lambda$ contributions.

\noi
At leading order in the ${1\over N_c}$ expansion, chiral symmetry is 
exact and mesons are stable and grouped in degenerated chiral multiplets.
At $O({1\over N_c})$ only the $\lambda $ term must be included. 
This term generates  (under the adequate ``conditions'') vacuum 
 instabilities in the 
LSM potential leading to the spontaneous breaking of chiral symmetry. New 
three-meson interactions and 
mass terms are generated. Higher order terms ($\beta$ and $\lambda^\prime$) 
allowed by chiral symmetry can then be incorporated. This procedure would 
give new mass terms coming from the $\lambda$ term only. The $U_A(1)$ 
breaking term would give just ($1/N^{3/2}_c$ suppressed) three-meson 
interactions. 
As we know that even before the chiral transition, both  the $U_A(1)$ 
symmetry is broken, and OZI rule violating transitions, though suppressed, 
are allowed transitions, 
we should include these terms before spontaneous symmetry breaking occur. 
Proceeding in this way, these terms also generates new mass terms and 
three-meson interactions which 
are formally subleading as compared with contributions coming from the 
$\lambda$ term. As the vacuum expectation values of 
scalar fields  are $O(\sqrt{N_c})$, the new mass terms and three meson 
interactions generated by the $U_A(1)$ breaking term are 
${\cal O}({1\over N_c})$ and ${\cal O}({1\over N^{3/2}_c})$ respectively. 
If we adopt the ${1\over N_c}$ counting rule for $\lambda^\prime$ as 
found in [25], 
meson masses contributions coming from the $\lambda^\prime$ term are ${\cal O}
({1\over N^{2}_c})$ while three-meson interactions generated by this term 
are ${\cal O}({1\over N^{5/2}_c})$. This is, however, not a clear issue as 
discussed in [25].

In the limit when ${c_0,c_8}$ (or equivalently the corresponding $c_s, c_{ns}$)
$\to 0$ we have spontaneous breaking of symmetry. In this case, and 
considering a  $SU(3)$ 
symmetric vacuum, pseudoscalar mesons are massless (GB) 
 except for the $\eta^\prime$ meson which acquire a non-zero
mass due to the combined effects of the $U_A(1)$-breaking and the spontaneous
breaking of symmetry. Notice however that this mass (related to t'Hooft mass 
in Ref. [25]) is formally subleading (${\cal O}({1\over N_c})$) in the 
${1\over N_c}$ expansion. In this case, the $\{|s>,|ns>\}$ pseudoscalar fields 
are mixed by the effects of the $U_A(1)$-breaking term.  The scalar meson 
octet acquire a common mass  
and the scalar singlet is split due to the effects of both, the $U_A(1)$- 
breaking term and the OZI rule forbidden (but $U(3)\times U(3)$ chirally 
symmetric) term $\lambda^\prime$.

\noi
Including the explicit chiral symmetry breaking terms $c_0, c_8$ (or 
equivalently $c_{ns}, c_{s}$) breaks the $SU(3)$ symmetry down to 
isospin, splits the isospin multiplets in both sectors and mixes the 
$\{ |1>,|8> \}$ fields. Pseudoscalar strange-non-strange fields remain mixed 
by the effects of the $U_A(1)$-breaking term only, while scalar 
strange-non-strange fields remain 
mixed by both, the $U_A(1)$-breaking term and the $\lambda^\prime$ term.

Let's now turn to the mixing analysis. For the final purposes of this work
(scalar mesons quark content and phenomenology) it is convenient to take the
$\{ |ns>,|s>\}$ basis in Eq.(13)as our starting point.
The physical fields are linear combinations of $\{ P_s, P_{ns}\}$ which 
diagonalize ${\cal L}^P_2$: 
\bq
\eta &=& P_{ns} cos(\phi_P) - P_s sin (\phi_P) \\
\eta^\prime &=& P_{ns} sin (\phi_P) + P_s cos (\phi_P). \nonumber
\eq
\noi The  relative angle between the $\{|1>,|8>\}$ basis  and the 
$ \{|S>,|NS>\}$ basis in Eq (13) corresponds to $\vartheta = -54.73^\circ$.

\noi The mixing angle can be obtained from the following relation
\be
sin ~(2\phi_P) = \frac{2 m^2_{P_{s-ns}}}{m^2_{\eta^{\prime}}-m^2_{\eta}} .
\en
The diagonal masses are
\bq
m^2_\eta &=& \frac{1}{2} (m^2_{P_{ns}} + m^2_{P_s}) - 
\frac{1}{2} R_P \nonumber \\
\mbox {} \\
m^2_{\eta^\prime} &=& \frac{1}{2}(m^2_{P_{ns}}+m^2_{P_s})+
\frac{1}{2}R_P\nonumber
\eq
\noi where
\be
R_P \equiv [(m^2_{P_{ns}}-m^2_{P_s})^2 +4m^4_{P_{s,ns}}]^{1/2}
\en
From the above relations trace invariance of the mass matrix is obvious
\be
m^2_{\eta^\prime}+m^2_{\eta}=m^2_{P_s}+m^2_{P_{ns}}.
\en

\noi
Similar relations to Eqs. (16,17,18,19) hold  for the corresponding scalar 
quantities with the $P\to S$ replacement. Using Eqs. (14,16)
we obtain

\be
sin (2\phi_P)= \frac{-4\sqrt{2} \beta a}
{m^2_{\eta^{\prime}}-m^2_{\eta}}
\en
\be
sin (2\phi_S)= \frac{4\sqrt{2} (\beta+\lambda^\prime b) a}
{m^2_{f^\prime}-m^2_f}
\en
\noi where we have denoted $f,f^\prime$ the physical scalar mesons analogous 
to the $\eta, \eta^\prime$ mesons.

From Eqs.(10,12) the Gell-Mann-Okubo (GMO) relations read
\bq
3 m^2_{8P} - 4 m^2_K + m^2_\pi&=&-2\lambda (a-b)^2 \\ 
3 m^2_{8S} - 4 m^2_\kappa + m^2_a&=&~2(\lambda + 2\lambda^\prime ) (a-b)^2.
\eq
\noi These equations reflect the well known fact that violations to GMO 
relations are second order in the $SU(3)$ breaking 
parameter. Notice that corrections to GMO relations have different sign in 
the scalar and pseudoscalar sectors.

On the other hand, Eqs.(17,18) yield 
\be
(m^2_{\eta^\prime} - m^2_\eta)^2 =
[\lambda (a^2-b^2)-2\beta b]^2 +32 \beta^2 a^2.  
\en
In addition to the $SU(3)$ breaking, the well known role of the $U_A(1)$ 
breaking in the $\eta, \eta^\prime$ mass 
splitting is explicitly exhibited in this last equation.

\bi
\bi
\noi
{\bf4. Predictions}

Before presenting  the phenomenological extraction for the 
 parameters entering in the model, we would like to stress that there are 
some predictions which are independent of some of the particular values for 
the parameters in the interacting Lagrangian. 
These predictions are relations between meson masses, three-meson couplings
and the scalar meson vacuum expectation values which are independent of the 
specific values for the remaining parameters entering in 
the model. Some of these relations which at present are interesting for 
phenomenological applications and can be straightforwardly deduced from 
${\cal L}_2,{\cal L}_3$ are:
\bq
 m^2_\kappa &=& {(a+b)m^2_K-2am^2_\pi \over b-a} \cr
 g_{a_0K^+K^-}&=& -{m^2_a-m^2_K \over \sqrt{2}(a+b)}\cr
 g_{a_0\kappa^+\kappa^-}&=& {m^2_a-m^2_\kappa \over \sqrt{2}(b-a)}\\
 g_{\pi_0\kappa^+K^-}&=&-{m^2_{\kappa}-m^2_\pi \over \sqrt{2}(a+b)}=
-{m^2_K-m^2_\pi \over \sqrt{2}(b-a)}\nonumber
\eq
The above relations concerns the non-mixed sectors. In these sectors, 
the OZI rule forbidden term $\lambda^\prime$ gives similar contributions 
to all masses (encoded in $\xi=\mu^2+\lambda^\prime (2a^2+b^2)$). This is 
not true for the mixed sectors where
the effects of the $U_A(1)$ breaking and OZI rule forbidden term 
$\lambda^\prime$ are important.
A more complete ( although not exhaustive) list of such relations arising 
solely from chiral symmetry and the way it is broken can be found 
in the general treatment of Schechter and Ueda[15]. As shown in that work, in 
a general treatment of the subject nothing can be said about the $I=0$ and 
$I=1$ scalar meson masses in the general case of chiral symmetry broken down 
to isospin.

There are two interesting relations more arising from LSM. The first one, 
concerns scalar and pseudoscalar mixing angles . In fact, from
 Eqs.(21,22) we obtain
\be
(m^2_{\eta^{\prime}}-m^2_{\eta})sin (2\phi_P)=-(m^2_{f^\prime}-m^2_f)
sin (2\phi_S) + 4\sqrt{2}\lambda^\prime ab.
\en 

\noi
The second, is a relation between scalar and pseudoscalar meson 
masses. From Eqs.(10,14,19 ) we get:
\be
m^2_{\eta^\prime}+m^2_{\eta}-2m^2_K=-(m^2_{f^\prime}+m^2_f-2m^2_{\kappa}-
2\lambda^\prime(2a^2+b^2))
\en
The $\lambda^\prime$ term in these equations is expected to be ${1\over N_c}$ 
suppressed as compared with the remaining terms [25] (the l.h.s in Eq. (27)
$= 2\beta (2a+b) + \lambda (a-b)= {\cal O}({1\over N_c})$), thus, disregarding 
this term LSM gives striking predictions for 
scalar meson masses and mixing angle $\phi_S$ which are valid modulo 
${1\over N_c}$ corrections.
In the case $\lambda^\prime = 0$, Eq.(27) reproduces Dmitrasinovic's sum rule 
[25].
 This sum rule was derived in the context of a 
Nambu-Jona-Lasinio model extended with a t'Hooft interaction which when 
bosonized reduces to a LSM like the one considered in this work, with definite
predictions for the parameters in Eq.(1). In particular, this approach 
predicts $\lambda^\prime =0$ [25]. We wish to emphasize, however, that 
the $\lambda^\prime$ term in Eq.(27) turns out to be crucial 
in the identification of scalar mesons. Without this term there seems to be 
no place for a light $\sigma$ as a member of the scalar meson nonet.

\bi
We now turn to the fixing of the parameters entering in the LSM.
The model has 6 free parameters $(a,b,\mu,\lambda^\prime,\lambda,\beta)$ which 
can be
fixed from phenomenology. As can be seen form Eqs (10,12,14), all the meson 
masses, 
but the mixed scalar meson ones, depend upon $(a,b,\lambda,\beta)$ and the 
combination $\xi =\mu^2+\lambda^\prime(2a^2+b^2)$. A phenomenological 
extraction of 
the value of $\lambda^\prime$ necessarily requires the use of information 
on the 
mixed scalar meson properties. However, this information is 
still 
controversial and the spirit of this work is to provide further understanding
on this sector. Instead of fixing the value of $\lambda^\prime$ we 
analyze 
scalar meson masses and mixing angle as a function of this parameter.

 By comparing Eqs. (5) with (9) and (10) we can see
that the pion and kaon decay constants are related to the scalar vacuum 
expectation values in the following way:
\be
f_\pi =\sqrt{2} a \qquad\qquad f_K = \frac{1}{\sqrt{2}} (a+b) .
\en

As the $\beta$ parameter is a direct measure of the $\eta_{NS} -\pi$ 
splitting we will use $(m_{\eta^\prime},m_\eta ,  m_\pi)$ in order to fix the 
values of $( \beta , \lambda ,\xi )$.
From Ecs(10,14,17,23) we obtain that $\beta$ is determined  by the following
equation:
\be
8(a^2+2b^2)\beta^2 - 4b(2m^2_\pi-m^2_{\eta^\prime}-m^2_\eta)\beta +
(m^2_{\eta^\prime}-m^2_\pi)(m^2_\eta-m^2_\pi)=0 
\en

\noi
and $\lambda$ and $\xi$ are obtained from:
\bq
\lambda &=&{2m^2_\pi- m^2_{\eta^\prime}-m^2_\eta -6b\beta \over a^2-b^2}\\
\xi &=& m^2_\pi-2b\beta-\lambda a^2 \nonumber
\eq
The solutions to these equations depend upon the scalar fields vacuum 
expectation values. We choose to fix $a$ only from Eq. (28) and analyze
everything as a function of $x={b-a\over2a}$ which is a measure of the
$SU(3)$ breaking.
The solutions to Eq.(29) are stable under changes in $x$.
This stability of the $\beta$ parameter is in contrast with the remaining 
parameters which are sensitive to the chosen value for $x$.
From the two solutions ($\beta\approx -750 MeV, -1550 MeV$) to  
equation (29),
the first one is ruled out by phenomenology. This can straightforwardly seen 
from the $K$ mass which is plotted in Fig. 1 as a function of $x$ for both 
solutions. The isospin averaged $K$ mass corresponds to 
$x\simeq 0.39$. However, as shown 
in Fig.1 the $K$ mass is not very sensitive to $x$ and allows for a wide range 
of values (roughly $x\in [0.3,0.5]$) for this parameter.

Concerning the sign of $\beta$ , it differ from the sign of the corresponding 
quantity obtained in Ref. [25].  The sign of $\beta$ 
can be reversed by redefining chiral transformations 
with a global sign. As a consequence, relations in Eq.(28) and solutions to Eq.
(29) gets modified by a sign, i.e. this change is equivalent to the 
transformations $\beta \to -\beta $ and $a,b \to -a,-b$. Meson masses and 
mixing angles are invariant under these transformations as can be easily seen 
from Eqs.(10,12,14). All three-meson couplings are
 changed by a global sign and four-meson couplings remain invariant. This 
relative sign between three and four-meson couplings should have  no physical 
consequences. This is so at least for meson decays and meson-meson scattering.
In other words, the $U_A(1)$ breaking term should not distinguish 
between otherwise equivalent degenerate vacuum states.

Meson masses for the non-mixed scalar 
sector ($a, \kappa$) do not depend on the 
particular value for $\lambda^\prime$ depending only on the $\xi$ combination 
of $\lambda^\prime$ and $\mu$. The mixed scalar meson 
($f, f^\prime$) masses do depend on the particular value for 
$\lambda^\prime$. 

\noi
In Fig.2 the dependences of $m_a, m_\kappa$ are shown as a 
function of  $x$ . The value $x=0.39$ corresponds to 
$m_\kappa\simeq 900 MeV, m_a\simeq 910 MeV$. A value of 980 MeV for the $a_0$ 
meson requires $x\simeq 0.27$ which is reasonably close to the set of values 
allowed by the fit to the $K$ mass.
  
\noi
Results for the mixed scalar meson masses are shown in Fig. 3 as a 
function of $\lambda^\prime$. 
Clearly the $f^\prime$ scalar  
 meson ( the scalar analogous to $\eta^\prime$ ) can be 
identified with the $f_0(980)$ meson  and the
 $f$ meson (scalar analogous to $\eta$) has a mass around 400
 MeV. Hereafter we call these mesons $f_0$ and $\sigma$ respectively. 
 A $980 MeV$ mass for the $f_0$ meson 
corresponds 
to $\lambda^\prime \approx 4$. This value gives $m_{\sigma}\approx 375 MeV$. 

Small values for $\lambda^\prime$ as required 
by the identification of the $f_0(980)$ as the scalar singlet suggested by 
Fig. 3 is 
roughly consistent with NJL model expectations [25] and Tornqvist 
conclusions (recall $\lambda^\prime = 4\lambda^\prime_T$)[9]. However, 
definitively LSM predicts a non-zero value for this parameter. On the other 
hand,
in Tornqvist picture the $\kappa$ meson is identified with the $K^*(1430)$ 
meson. A $\kappa$ meson mass of 1430 MeV would require $x\simeq 0.14$ which 
is unlikely in 
the light of the the results for the $K$ mass shown in Fig. 1. Furthermore, 
a calculation of the 
$a_0\to \gamma \gamma$ decay within a SU(3) Linear Chiral Model gives 
support to the identification of both the $a_0(980)$ and the $\kappa
(900)$ as members of the scalar $\bar q q$ nonet [26]. A complete reanalysis 
of scalar meson interactions within LSM is under investigation.

As to the mixing angles, the pseudoscalar mixing angle (which is independent 
of  $\lambda^\prime$) , can be extracted from Eq.(20). We obtain
 $ 33.2^\circ\le\phi_P\le 38.5^\circ$  for $0.27\le x\le 0.39$. 
This range of values correspond to $ -16.2^\circ\le
\theta_P\le -21.5^\circ$ for the pseudoscalar mixing angle in the usual 
singlet-octet basis. The central value ($ x=0.335$) $\phi_P=35.5^\circ$ is in 
agreement with the
experimental result as extracted from $\eta, \eta^\prime$
radiative decays($\theta_P \simeq -19^\circ$)[8,27]. This result for the 
pseudoscalar mixing angle can be considered as a consistency check for the 
model.

\noi
The scalar mixing angle can be extracted from  
Eq.(21). This angle depends on $\lambda^\prime$ and for 
say, $\lambda^\prime=4$ and $x=0.39$ we 
obtain $\phi_S \approx -14^\circ$. This value is very stable  under changes
in $x$. It roughly lies in the range $-5^\circ \le \phi_S\le -19^\circ$ for 
$10\ge \lambda^\prime\ge 0$. 

Concerning the kaon decay constant in Eq.(28), the predicted value  is 
$f_K=(1+x)f_\pi$ with $x\in [0.27, 0.39]$. This value  is not too far from 
its experimental result [8] $f^{exp}_K=1.22 f^{exp}_\pi$.
 Furthermore, one-loop corrections[15]  to $x$ have 
been shown to be $\simeq 10\%$ ($x$ corresponds to $\simeq 2 $ times the 
expansion parameter
${\{\sigma_8\} \over \sqrt{2}\{\sigma_0\}}$ used by Chan and Haymaker in 
reference [15]) and in the 
right direction to bring down the predicted value for $f_K$  even closer to 
experimental results. A complete reanalysis of the results of the model at 
one one-loop level would be desirable to confirm the gross results emerging 
at tree level.

\bi  
\noi
{\bf 5. CHPT low energy constants and LSM}

As well known, ${\cal O}(p^4)$ low energy constants can be fixed by invoking 
saturation by meson resonance exchange. In particular, the lowest order 
lagrangian involving scalar meson resonances is [11]:
\be
{\cal L}_2[ S(0^{++})]= c_d\bigl< S u_\mu u^\mu \bigr> + c_m\bigl< S \chi_{+}
\bigr> + \tilde c_d S_1\bigl< u_\mu u^\mu \bigr> + 
\tilde c_m S_1\bigl< \chi_{+}\bigr>
\en
where
\be
u_\mu = iu^\dagger D_\mu U u^\dagger =u^\dagger_\mu, ~~~ 
U=u^2=exp(-i\sqrt{2}\Phi/F),~~~\Phi={1\over \sqrt{2}}\sum_{i=1}^8\lambda_iP_i
\en

with
\be
\chi_\pm= u^\dagger \chi u^\dagger \pm u\chi u, ~~~ \chi=2B_0{\cal M}.
\en
Clearly, Eq.(31) do not consider scalar meson mixing which arise at higher 
order in the chiral expansion. 
As discussed in the first of Refs. [11], scalar meson resonance exchange gives 
the following contributions to the ${\cal O}(p^4)$ low energy constants:

\noi
Octet contributions:
\be
L^S_1=-{c^2_d\over 6M^2_S},~~~L^S_3={c^2_d\over 2M^2_S},~~~~
L^S_4=-{c_dc_m\over 3M^2_S},~~~L^S_5={c_dc_m\over M^2_S}
\en
\be
L^S_6=-{c^2_m\over 6M^2_S},~~~L^S_8={c^2_m\over 2M^2_S},~~~
H^S_2={c^2_m\over M^2_S}.
\en
Singlet contributions:
\be
L^{S_1}_1={\tilde c^2_d\over 2M^2_{S_1}},~~~
L^{S_1}_4={\tilde c_d \tilde c_m\over M^2_{S_1}},~~~
L^{S_1}_6={\tilde c^2_m\over 2M^2_{S_1}}
\en

\noi
Expanding Eq. (31) we obtain
\bq
{\cal L}_2[ S(0^{++})]&=& {2c_d\over F^2}\bigl< S \partial_\mu \Phi 
\partial^\mu \Phi\bigr> + {2\tilde c_d\over F^2} S_1\bigl< \partial_\mu \Phi 
\partial^\mu \Phi\bigr>+ \\ \nonumber
 & & 4B_0c_m\bigl[\bigl< S {\cal M}\bigr> 
-{1\over 4F^2}\bigl<S(\Phi^2{\cal M}+ {\cal M}\Phi^2+ 2\Phi{\cal M}\Phi)\bigr>
\bigr] + \\ \nonumber
& &4B_0\tilde c_m S_1\bigl[\bigl< {\cal M}\bigr> 
-{1\over 4F^2}\bigl<\Phi^2{\cal M}+{\cal M}\Phi^2+2\Phi{\cal M}\Phi \bigr>
\bigr] +
{\cal O}(\Phi^4)
\eq
\noi
where F denotes the pseudoscalar decay constant in the chiral limit. 

\noi
We can now make contact with LSM results in the SU(3) symmetric limit $a=b$.
In this limit  a comparison of LSM with Eq.(37) gives:

\be
c_d=- 2c_m=-{F\over \sqrt{2}},~~~\tilde c_d=-2\tilde c_m=-{F\over \sqrt{6}}.
\en
\noi 
As expected [11] the couplings $c_d,~c_m,~\tilde c_d,~\tilde c_m$ are 
${\cal O}(N^{{1\over 2}}_c)$ and satisfy the large $ N_c$ relations
\be 
\tilde c_d={\varepsilon\over \sqrt{3}} c_d, ~~~
\tilde c_m={\varepsilon\over \sqrt{3}} c_m,~~~~\varepsilon= 1.
\en

\noi
These results gives the following contributions of scalar resonances exchange 
to the ${\cal O} (p^4)$ low energy constants

\bq
L^{SC}_1&=&~~{F^2\over 12}({1\over M^2_{S_1}}-{1\over M^2_S}),~~
L^{SC}_3=~~{F^2\over 4 M^2_S}, \\ \nonumber
L^{SC}_4&=&-{F^2\over 12}({1\over M^2_{S_1}}-{1\over M^2_S}),~~
L^{SC}_5=-{F^2\over 4 M^2_S}, \\ \nonumber
L^{SC}_6&=&~{F^2\over 48}({1\over M^2_{S_1}}-{1\over M^2_S}),~~~
L^{SC}_8={F^2\over 16 M^2_S},~~~
H^{SC}_2={F^2\over 8 M^2_S}. \\ \nonumber
\eq

There is a worth remarking point about Eq.(40).  From Eqs. ( 14), 
in the $SU(3)$ symmetric limit we obtain :

\bq
m^2_{0P}-m^2_{8P}&=&-6\beta a \\ \nonumber
m^2_{0S}-m^2_{8S}&=&6(\beta+\lambda^\prime a) a. \nonumber 
\eq

\noi
From the above relations is clear that the $U_A(1)$ breaking 
has an inverted effect in the mixed scalar sector as compared with its 
effect in 
the pseudoscalar sector. In the pseudoscalar 
sector the anomaly pushes the singlet up and the octet down while in the 
scalar sector it does push the singlet down and the octet up. This effect 
changes the signs of $L_1,~L_4,~L_6$ from the naively expected ones as 
$m^2_{0S}$ turns out to be smaller than $m^2_{8S}$ due to the $U_A(1)$ 
breaking effect in the scalar sector.

Scalar mesons exchange contributions to the LEC quoted in Eq. (40) have the 
right large $N_c$ properties. In fact, from
 Eqs. (40,41) is clear that the contribution 
of scalar mesons exchange to $L^{SC}_1$ is formally ${\cal O}(1)$ as it must 
be since, although $L_1$ is ${\cal O}(N)$, $2L_1-L_2$ is ${\cal O}(1)$ and 
there are no scalar contributions to $L_2$. Likewise, the remaining LEC in 
Eq. (40) can 
be easily seen to have  the right large $N_c$ behavior.

The scalar couplings in Eq.(31) can be estimated from (38,39). Using $F=f_\pi 
=93~MeV$ we obtain:
\bq
c_d&=&-6.56\times 10^{-2}~GeV,~~~~~c_m=3.28\times 10^{-2}~GeV \\ \nonumber
\tilde c_d&=&-3.79\times 10^{-2}~GeV~~~~~\tilde c_m=1.89\times 10^{-2}~GeV \\ 
\nonumber
\eq
Notice that LSM predicts $c_d c_m <0$, thus the expected saturation of $L_5$ 
by scalar meson exchange [11] is not satisfied. This can be explicitly seen by 
numerically evaluating the scalar contributions to CHPT LEC in Eq.(40). A  
first estimate can be obtained by using the large $N_c$ limit 
where $M_S=M_{S_1}$. In this limit, the only non-zero LEC are
\be
L^{SC}_3=2.24\times 10^{-3},~~~L^{SC}_5=-2.24\times 10^{-3},~~~
L^{SC}_8=0.56\times 10^{-3}.
\en
We have used $M_S=M_{a_0}=980~MeV$ in obtaining these values. The value of 
$L^{SC}_8$ above is consistent with the saturation by scalar mesons exchange
invoked in Ref.[11]. A calculation of $L^{SC}_8$ using the LSM estimate for the 
averaged octet mass ($M^{LSM}_S\simeq 912~MeV$) gives 
$L^{SC}_8=0.64\times 10^{-3}$ to be compared 
with the experimental value $L^{exp}_8(m_\rho )=0.9\pm 0.3\times 10^{-3}$ [11].
These results are consistent with the expected saturation of $L_8$ by scalar 
mesons exchange. However, $L^{SC}_5$ has the wrong sign and LSM 
predicts that this LEC is not saturated by scalar mesons exchange. As to 
$L^{SC}_3$ it has the same sign as that found in [11] but  LSM predicts a 
stronger effect than the one quoted there, thus reducing the total value of 
the resonance exchange contribution to this LEC quoted in Table 3 of Ref. [11].

\bi

{\bf 6. Conclusions.}

We work out $U(3)\times U(3)$ Linear Sigma Model predictions  
for meson masses and mixing angles at tree level. We clearly exhibit the 
effects of the $U_A(1)$-breaking term in the generation of meson 
masses and in the mixing of the scalar and pseudoscalar mesons.
The parameters 
entering in the model are fitted to the pseudoscalar spectrum. The model 
predicts that the members of the  scalar meson nonet are:  
$\{ a_0(980), \kappa(900),\sigma(\approx 400), f_0(980)\}$ 
resonances. These mesons are the analogous to $\{\pi,K,\eta,\eta^\prime \}$ 
pseudoscalar mesons. The scalar mixing angle in 
the $\{ |ns>, |s>\}$ basis is 
found to be $\phi_S\approx -14^\circ$, thus the $f_0(980)$ meson is mostly 
strange and the $\sigma(\approx 400)$ is predominantly non-strange. 
The pseudoscalar mixing angle in this basis is $\phi_P\approx 35^\circ$ which 
corresponds to $\theta_P\approx -19^\circ$ in the singlet-octet basis.
The fit of the parameters of the model gives $\lambda^\prime \approx 4$. This 
result is along the line of 
Tornqvist's conclusion in the sense that the $\lambda^\prime_T$ 
parameter(${\lambda^\prime
\over 4}$ in this work) is small [9]. However, we 
obtain a $\kappa$ meson mass $\approx 900 MeV$ in accordance to Ishida [20] 
reanalysis of $K \bar K -\pi\pi$ scattering data,
thus being unlikely to identify this particle with $K^*(1430)$ meson in this 
model. 
We also study the contributions of scalar mesons exchange to the 
${\cal O}(p^4)$ low energy constants of CHPT. The model predicts saturation of
$L_8$ by scalar resonances exchange as expected [11]. However, $L_5$ is not 
saturated by these resonances.

\bi
\noi
{\bf Acknowledgments.}

I wish to acknowledge the hospitality of the high energy physics theory group
at UCSD where part of this work was done. Special thanks are given to V. 
Dmitrasinovic for calling my attention to his work. This work was supported in 
part by Conacyt, Mexico, under project 3979PE-9608 and grant 933053.

\bi
\vfill
\eject

{\bf References.}
\begin {itemize}
\item[1.-] S. Weinberg Physica A96 327 (1979);J. Gasser and L. Leutwyler 
Nucl. Phys. B250, 465 (1985);J. Gasser , M.E. Sainio and V. Svarc Nucl. 
Phys. B307, 779 (1988). For a review see e.g.``Dynamics of the Standard 
Model'' J.F. Donoghue,E. Golowich and B.R. Holstein Cambridge Univ. 
Press (1992).
\item[2.-] E. Jenkins and A. V. Manohar Phys. Lett. B255, 558 (1991); 
B259, 363 (1991); 
E. Jenkins, A.V. Manohar and M. Wise Phys. Rev. Lett. 75, 2272 (1995).
\item[3.-] T. R. Hemmert, B. R. Holstein and J. Kambor hep-ph/9712496 
(unpublished)
\item[4.-] H. Leutwyler Phys. Lett B374; P. Herrera-Siklody et. al \\
hep-ph/9610549.
\item[5.-] G. t'Hooft, Nucl. Phys B72, 461 (1974) ; E. Witten. Nucl. Phys.B160,
 57 (1979)  ; R. Dashen, E. Jenkins and A. Manohar Phys Rev D49, 4713 (1994); 
D51, 3697 (1995). 
\item[6.-] see e.g. H. Marksiske et.al. The Crystal Ball Coll. Phys. Rev. D41, 
3324 (1990); T. Oest et. al. JADE Coll. Z. Phys. C47, 343 (1990)  and 
references therein. 
\item[7.-] T. Barnes Phys. Lett. B165, 434 (1985); 
\item[8.-] Particle Data Group, R. M. Barnett et. al.  Phys. Rev. D54, 1 
(1996).
\item[9.-] N.A. Tornqvist hep-ph/9711483,hep-ph/9712479, unpublished; 
N.A. Tornqvist and M.Ross, Phys. Rev. Lett. 76 (1996) 1575. 
\item[10.-]  F. Sannino and J. Schechter, Phys. Rev. D52:96 (1995); M. Harada
F. Sannino and J. Schechter, Phys. Rev. D54, 1991 (1996); Phys. Rev. Lett. 78 
,1603 (1997). 
 \item[11.-] G. Ecker et.al. Nucl. Phys. B321, 311, (1989); J. F. Donoghue, 
C. Ramirez and G. Valencia, Phys. Rev. D39, 1947 (1989).
 \item[12.-] M. Napsuciale and J. L. Lucio Phys. Lett. B384, 227 (1996) ; 
Nucl. Phys. B494, 260 (1997).
\item[13.-] V. Pascalutsa hep-ph/9802282 (unpublished).
\item[14.-] J. Binjnens and F. Cornet, Nucl. Phys. B296, 557 (1988); \\
J. F. Donoghue, B.R.Holstein and Y.C.Lin, Phys. Rev. D37, 2423 (1988); 
S. Bellucci,J. Gasser and M.E. Sainio, Nucl. Phys. B423, 80 (1994).
\item[15.-] J. Schechter and Y. Ueda Phys. Rev. D3 2874 (1971); L.H. Chan 
and R.W. Haymaker Phys. Rev. D7 402 (1973); Phys. Rev. D7 415 (1973).
\item[16.-] V. De Alfaro et.al. ``Currents in hadron Physics'', 
North Holland Publ.( 1973) Amsterdam, Chap. 5.
\item[17.-] C. J. Morningstar and M. Peardon Phys. Rev. D56, 4043 (1997);
M. Teper hep-ph/9712504, unpublished; V. Anisovich, hep-ph/9712504, 
unpublished.
\item[18.-] L.R. Baboukhadia, V. Elias and M.D. Scadron hep-ph/9708431, 
unpublished.
\item[19.-] A. Bramon, R. Riazuddin and M.D. Scadron hep-ph/9709274, 
unpublished.
\item[20.-] S. Ishida et.al. hep-ph/9712230; M.Y. Ishida and S.Ishida \\
hep-ph/9712231 to be published in Proc. of Int. Conf. Hadron'97.
\item[21.-] M.Levy Nuov. Cim. LIIA 7247 (1967).
\item[22.-] S. Gasiorowicz and D.A. Geffen Rev. Mod. Phys. 41 531 (1969).
\item[23.-] G. t'Hooft, Phys. Rep. 142, 357, (1986).
\item[24.-] M. D. Scadron Phys. Rev. D26 239 (1982).
\item[25.-] V. Dmitrasinovic. Phys. Rev C53, 1383, (1996).
\item[26.-] M. Napsuciale, J.L.Lucio and M.D. Scadron in preparation.
\item[27.-] E.P.Venugopal and B.R.Holstein hep-ph/9710382, unpublished.

\end {itemize}
\vfill
\eject

\noi
{\bf Figure caption}

\begin{itemize}
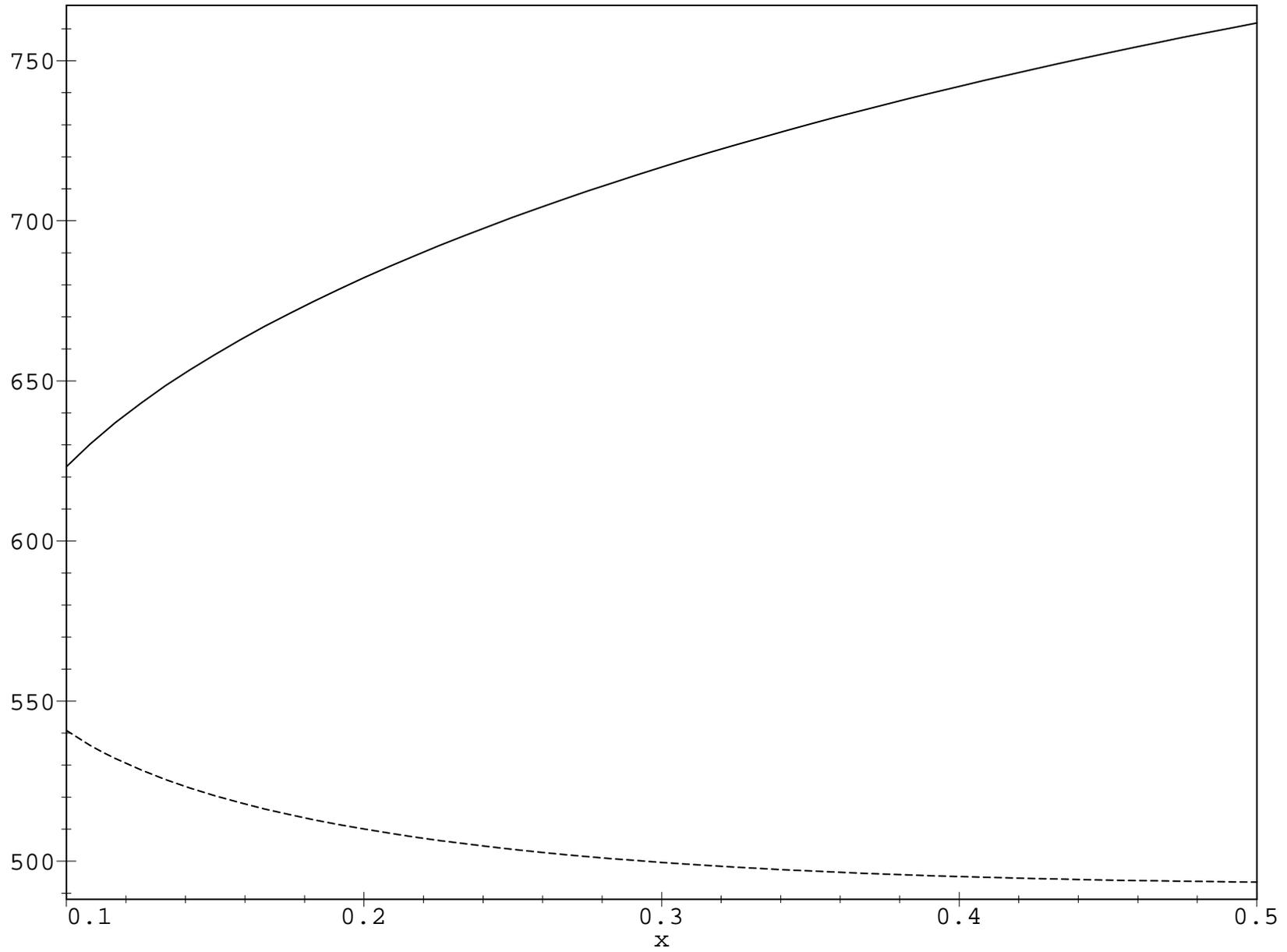
\item[Fig.1.-] $K$ meson mass (in units of $MeV$) as a function of $x={b-a\over 2a}$ for the two 
solutions to Eq.(29) $\beta \approx -700 MeV$ (continuous line) 
and $\beta\approx -1550 MeV$(dashed line).

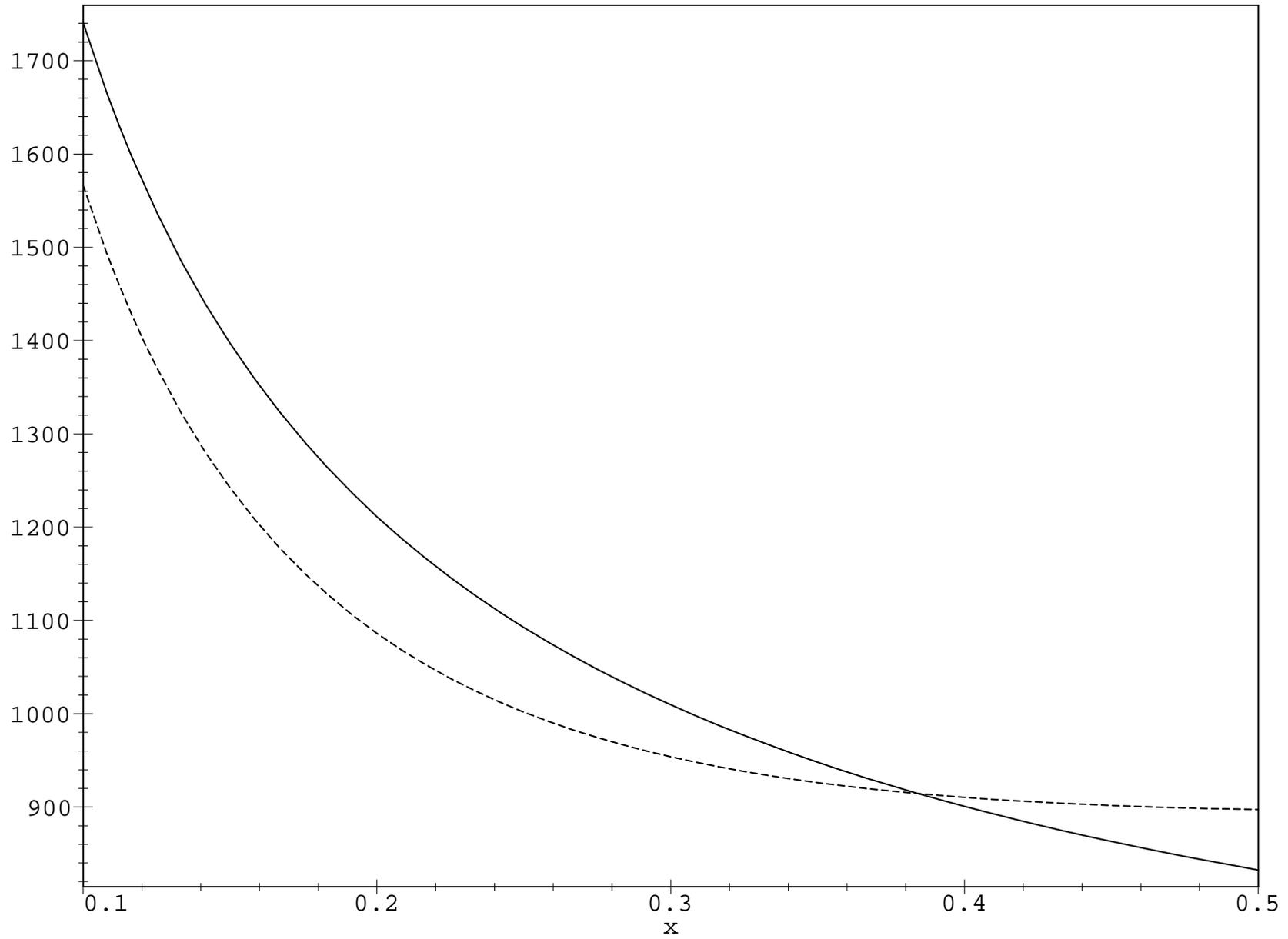
\item[Fig.2.-] $a_0$ (dashed line) and $\kappa$ (continuous line) meson masses
(in units of $MeV$) as a function of $x={b-a\over 2a}$.

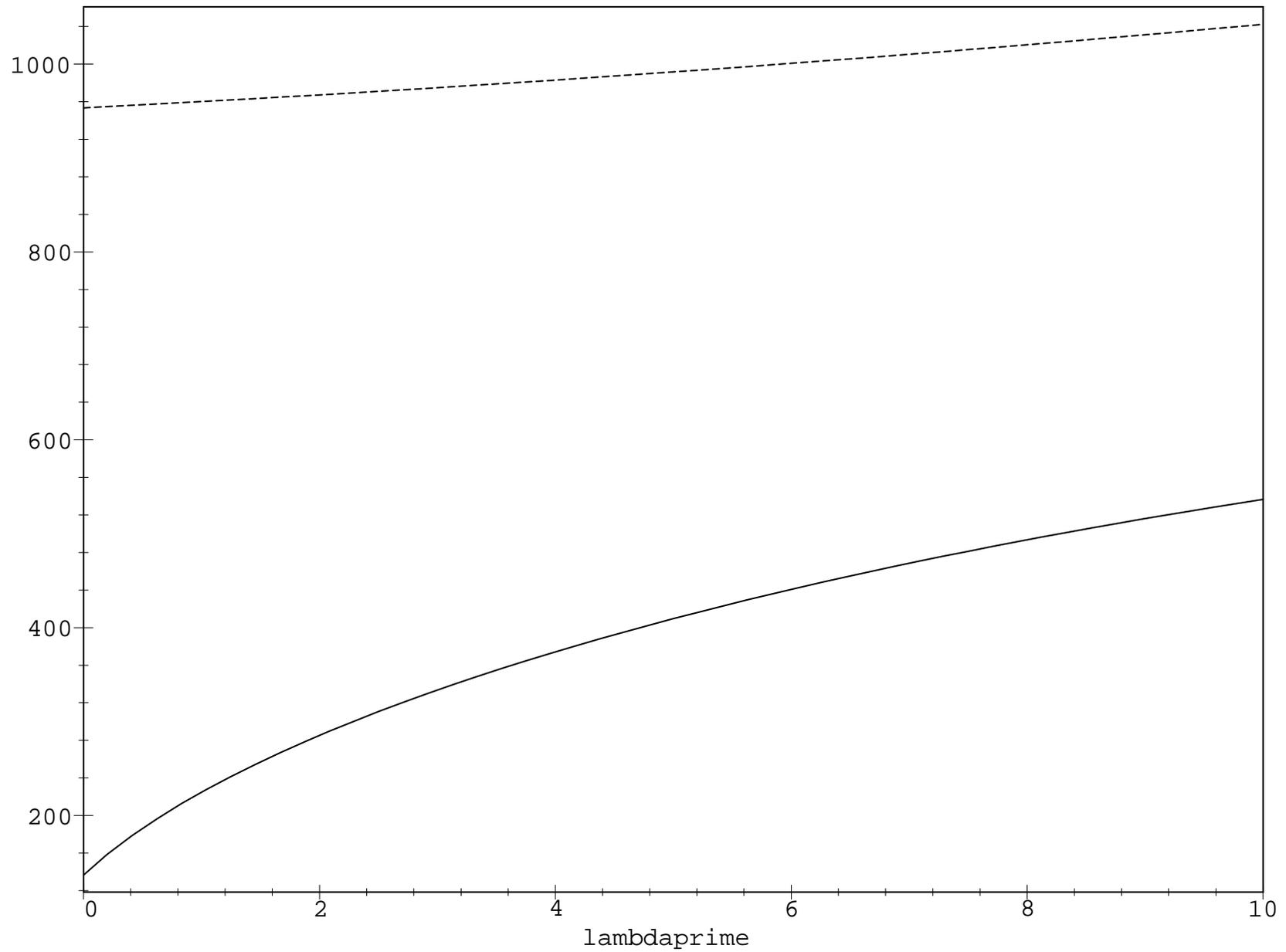
\item[Fig.3.-] $\sigma$ (continuous line) and $\sigma^\prime$ (dashed line) 
masses (in units of $MeV$) as a function of $\lambda^\prime$ for $x=0.39$.

\end{itemize}

\end{document}